# Design of Image Cryptosystem by Simultaneous VQ-Compression and Shuffling of Codebook and Index Matrix

Arup Kumar Pal, G.P. Biswas and S. Mukhopadhyay

Department of Computer Engineering, Indian School of Mines,
Dhanbad-826004, India

arupkrpal@gmail.com, gp_biswas@yahoo.co.in and msushanta2001@yahoo.com

## Abstract

*The popularity of Internet usage although increases exponentially, it is incapable of providing the security for exchange of confidential data between the users. As a result, several cryptosystems for encryption of data and images have been developed for secured transmission over Internet. In this work, a scheme for Image encryption/decryption based on Vector Quantization (VQ) has been proposed that concurrently encodes the images for compression and shuffles the codebook and the index matrix using pseudorandom sequences for encryption. The processing time of the proposed scheme is much less than the other cryptosystems, because it does not use any traditional cryptographic operations, and instead it performs swapping between the contents of the codebook with respect to a random sequence, which resulted an indirect shuffling of the contents of the index matrix. It may be noted that the security of the proposed cryptosystem depends on the generation and the exchange of the random sequences used. Since the generation of truly random sequences are not practically feasible, we simulate the proposed scheme using MATLAB, where its operators like **rand(method, seed), randperm(n)** has been used to generate pseudorandom sequences and it has been seen that the proposed cryptosystem shows the expected performance.*

## 1. Introduction

With the rapid growth of usage of multimedia data, the demand for new approaches of their secured transmission over public channel is also rapidly increasing. This has opened opportunity for extensive research for cryptosystem based secured transmission of multimedia data. Digital image compression [1] and encryption technique [2] will be the solution to solve the problems stated above. While compression technique reduces the size of an image file to be transferred or stored, encryption is widely used to ensure security. To achieve the goal, if some existing compression and encryption techniques are combined together, it leads to another problem. The overall processing performance will be slowed due to many computational involvements in these two techniques. Furthermore, one essential point before designing the image cryptosystem is that image must be compressed previous to encryption operation. The reason is that the compression of the encrypted images contributes to low compression ratio (CR) than the compression of unencrypted images as the redundancy of original images is broken on encryption. For this, a suitable and fast cryptography technique has been adopted on compressed image data. The other issue for designing image cryptosystem is that traditional data encryption techniques are not applicable on image data. An image requires more time during encryption/decryption process when a traditional data encryption/decryption techniques are used, because image data is huge in size. Besides, when the traditional data encryption is used on text data, the decrypted text must be identical to the plaintext. However, this requirement is not always necessary for image data. The decrypted image need not be faithful replication of the original image because human visual perception allows the loss of visual intimation to a considerable extent. Because of these differences between image data and text data as well as compressing data speeds-up the encryption/decryption process, in literature several image cryptosystems have been proposed on compressed image. For example, C-C Chang et al. [3] first applied vector quantization to compress the image. After that for enhancing security they diffused and confussed the codebook and encrypted these parameters of the codebook by a symmetric cryptosystem like DES [2]. Tung-Shou et. al. [4] proposed scheme is known as virtual image cryptosystem. In their technique, the VQ-based secret image is embedded into the host image to form the stego-image. In [5], Chang et al. proposed scheme can achieve lossless compression using quadtree data structure as well as appreciate secure encoding through secret scanning sequences. In Chuanfeng et al. [6], proposed image cryptosystem is based on k-PCA based compressed image data. Then the compressed image is fed to RC4 algorithm for encryption. This method can achieve a high security against different kinds of attacks. Maniccam et. al. [7] proposed scheme is based on different scan patterns, these patterns are used to compress the image in lossless fashion then the compressed image is encrypted by rearranging the bits of the compressed image using a set of scanning paths. In [8], Goh et. al. used DWT and EZW to compress the input secret image. After the compression, RC4 algorithm was applied on the compressed image to ensure confidentiality of image. But this technique does not provide any compression after applying encryption algorithm on compressed image. Among them, Vector Quantization (VQ) based image cryptosystems are attractive especially because it is a block-based quantization method with the benefits including simple architecture, efficient decoding process, and easy implementation.

In this paper, we have proposed an image cryptosystem based on the VQ scheme. The conventional VQ-scheme compresses the input image to a codebook and an index-matrix. Then the codebook and the index-matrix are encrypted for secure transmission over public channel. Our proposed method can serve the following objectives: (i) we can achieve both the compression and the encryption of the input image simultaneously; (ii) the proposed method can be easily implemented and highly efficient for the fast encryption and decryption of the compressed image; (iii) the encryption mechanism makes the encrypted data more secure without using popular traditional cryptography technique such as DES, RSA and so on.

The rest of this paper is organized as follows: In section 2 we present a brief overview of VQ based image compression and subsequently several approaches for designing of VQ based image cryptosystem. The proposed image cryptosystem is described in section 3. The experimental result and security analysis of the proposed image cryptosystem are presented in section 4 and 5 respectively. Finally, the conclusion is given in section 6.

## 2. PRELIMINARIES OF VQ AND DESIGN OF VQ BASED IMAGE CRYPTOSYSTEM

The proposed image cryptosystem, during VQ encoding time, simultaneously performs the compression and encryption of the input image for transmission over insecure communication channel. For the sake of completeness, this section gives some basics of Vector Quantization for image compression. In addition, several existence approaches for designing VQ based image cryptosystem are described with their schematic diagram.

### 2.1. VQ Based Image Compression

The VQ [9-12] is one of the popular lossy image compression techniques and it is based on the basic idea, according to the Shannon's rate-distortion theory, that the better compression is always achievable by coding image vectors not image scalars. It provides two advantages of reducing the required bit-rate for image transmission and requiring simple hardware structure for image decoding at the receiving end. Initially, the VQ encodes original images by decomposing into a set of vectors, and then it uses a suitable codebook to match and represent those vectors in the original images, where the size of each codeword of the codebook can be typically $4 \times 4$ pixels. The matching is obtained based on the smallest squared Euclidean distances between the image vectors and the codeword of the codebook; the indices of the matched codeword are then used to replace the image vectors for compression. In case of decoding, each index is used to search the same codebook and the corresponding codeword is placed in the position indicated by the index to get decompressed images. The algorithmic steps of VQ are given.

**Algorithmic steps of VQ Compression**

**Step 1:** Initially, an input image X is divided into non-overlapping *n* blocks of typical size say *4×4* pixels. Then these blocks are converted into vectors of dimension *16* such that X={$X_1$, $X_2$, $X_3$,…, $X_{n-1}$, $X_n$}, where $X_i$ is the i-th training vector.

**Step 2:** A suitable codebook of m codewords say, Y = {$Y_1$, $Y_2$, $Y_3$,…, $Y_{m-1}$, $Y_m$} is considered by using LBG [11-12] algorithm, where *n>m*. Then the image vectors formed in step-1 are scanned serially and they are represented with an index i corresponding to the elements $Y_i$, *for i = 1 to n*, in Y, that closely match with the input image vectors. The matching is done using a distance metric such as Euclidean distance technique.

**Step 3:** The indices and the selected codewords obtained in step-2 are transmitted separately to the destination, where they are decoded for the construction of original images as shown in *Fig. 1.*

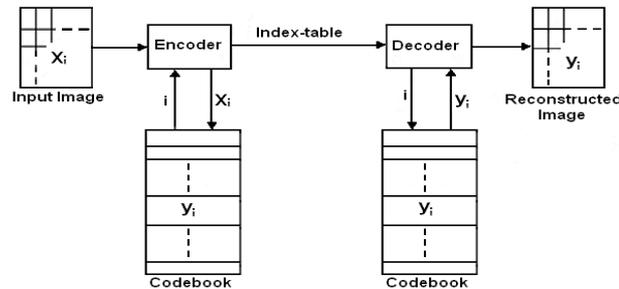

*Fig. 1.* The block diagram of encoding and decoding in VQ

**2.2. VQ Based Image Cryptosystem Design**

As stated earlier, conventional VQ compresses the input image into the combination of the codebook and the index-matrix. Now the codebook and/or the index-matrix are encrypted to make them confidential. This can be achieved in several ways and among them three major techniques which are adopted by different researcher are described in this section.

**Mode 1:** The codebook is encrypted before sending to the receiver. The index-matrix is sent as a plain text form. The schematic diagram of such image cryptosystem is shown in *Fig 2.*

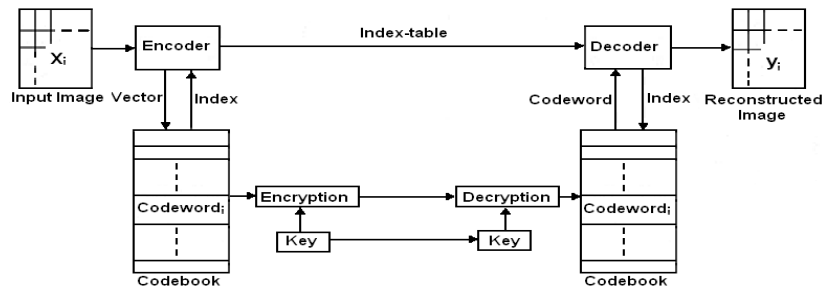

*Fig. 2.* The schematic diagram for enciphering the codebook in VQ

**Mode 2:** In this mode the index-matrix is selected for encryption process and the codebook is transmitted through public channel without encryption. The block diagram of such image cryptosystem, where index-matrix is considered for encryption is shown in *Fig. 3*.

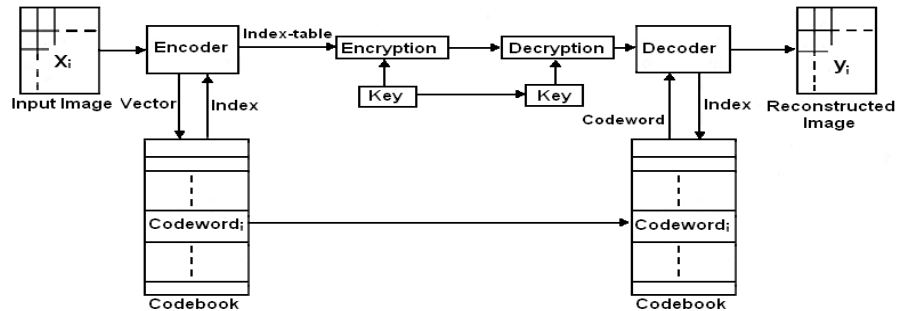

*Fig. 3.* The schematic diagram for enciphering the index-matrix in VQ

**Mode 3:** Both the compressed data set i.e. codebook and index-matrix is encrypted to ensure its security. This scheme is much more secure compare to the first two modes. The schematic diagram of such image cryptosystem is shown in *Fig. 4*.

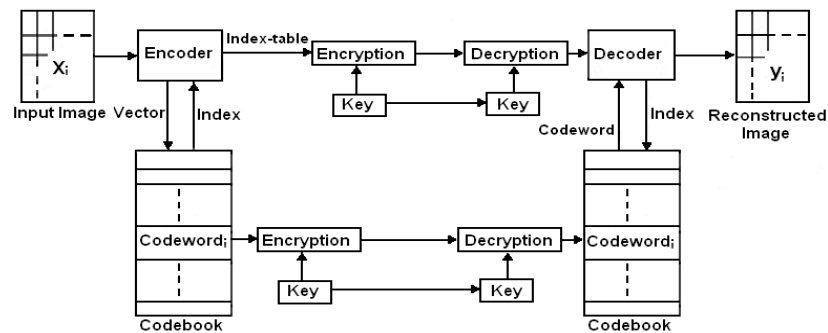

*Fig. 4.* The schematic diagram for enciphering both the codebook
and index-matrix in VQ

In this paper, the proposed technique basically compresses and encrypts the input image during VQ encoding time. Also the proposed technique, which provides high degree of security, instead of using any conventional cryptography technique and its execution time is fast, is described in the subsequent section in details.

## 3. THE PROPOSED SCHEME

The detail description of the proposed scheme for image compression and encryption are given in this section. The objective of the proposed technique is that if the contents of codebook and/or index-matrix are shuffled before transmitting, then the decryption is not possible without knowing the reverse shuffling process. The proposed technique starts with generating a set of random numbers and each of which is assigned to each codeword sequentially. Then during the VQ encoding time, it generates an index-matrix, whose entries are random index of codeword. If this generated index-matrix and codebook are transmitted to the receiver, then the receiver is not able to reconstruct the original image without knowing the sequence of random numbers. However in some cases, such type of image cryptosystem does not provide satisfactory result. This happens due to the characteristic of image, in which cases the image contains several flat regions. These flat regions are generally encoded by a single codeword and corresponding random index is recorded into index-matrix. So in decoding time, when the same random index sequence is not assigned to the codebook, then the codebook is searched sequentially with respect to each received random index, and the corresponding image block is placed over the received index. In such way, the reconstructed image contents the same flat region with different codeword. To overcome this problem, the proposed technique has modified the VQ encoding-decoding process. In the proposed technique, after encoding the input vector with a codeword in a codebook, the encoded codeword would be swapped with respect to a codeword, which is available sequentially in the codebook at the position value same to the encoded codeword's random index value.

To explain this in details let us consider an image of size *20×20* pixels. The image is divided into blocks of sub image of size *4×4* pixels and a codebook is constructed having four codewords. Then codewords, which are denoted as C1, C2, C3, C4 are indexed as 4,1,3,2 respectively. It implies that, the random index position of first codeword, C1 is represented as 4 and so on. If we consider that all image blocks are encoded into codeword, C4 then the generated index-matrix would be as follow: for the case of first image block, the entries of index-matrix would be 2 since the encoded codeword is C4 and after that it performs swapping between C4 and C2 codeword. So the codebook sequence would be C1,C4,C3,C2 but its random sequence remains unchanged. Similarly after encoding the second image block with C4, it stores 1 in the index-matrix followed by swapping operation. This process would be continued until the last image block is processed. In case of last image blocks, the swapping operation is not performed. The generated index-matrix is shown in *Fig. 5*.

| 2 | 1 | 4 | 2 | 1 |
|---|---|---|---|---|
| 4 | 2 | 1 | 4 | 2 |
| 1 | 4 | 2 | 1 | 4 |
| 2 | 1 | 4 | 2 | 1 |
| 4 | 2 | 1 | 4 | 2 |

*Fig. 5.* Index-matrix

Although, all the image block are mapped into C4 but in the above index-matrix, index of C4 are changed during each image block encoding time. Thus after VQ encoding process it produces a shuffled index-matrix and with random indexed codebook. Thus image compression-encryption is possible jointly. The pseudo code for the image encryption-decryption of the proposed method is given below.

### 3.1    Encryption and Decryption of Images

As stated earlier, the objective of the proposed technique is to generate a shuffled index-matrix and codebook, which are done in this work. Here shuffling is consider as encryption process. Shuffling (reverse shuffling) process is happened during VQ-encoding (decoding) time. The encoding and the decoding processes comprise of number of operations as given below, where the decoding starts from last index of the index-matrix.

**Encoding**

1. The LBG algorithm is applied to the given input image to generate the codebook.
2. A pseudorandom number generator (PRNG) and a secret seed are selected for the above image cryptosystem; those are then used to generate a pseudorandom number sequence. The generated sequence contains random numbers equal to the number of codeword in the codebook.
3. The generated random numbers are used to assign each codeword sequentially as an index.
4. Image is decomposed into non-overlapping blocks and each block is matched/searched with the codebook generated in (1) as explained below.

   *for i =1 to no_block*
   *original_index:= encode(codebook, block[i])*
   *pseudo_index:= random_number[original_index]*
   *index_table [i]:= pseudo_index*
   *if ( i < no_block )*
   *swap(codebook[original_index], codebook[pseudo_index])*
   *end for*

**Decoding**

1. On receiving, the seed through secure channel/key exchange technique, the same sequence of pseudorandom numbers is generated as used in case of encoding process, and each of which is assigned to each codeword sequentially as an index.
2. The codebook is then searched with respect to received index-matrix as follows:

   *for j= 1 to no_codeword*
   *original_index:=position(random_number[j]= =index_table[no_block])*
   *end for*
   *swap(codebook[index_table[no_block]],codebook[original_index])*

   *for i = no_block to 1*
   *block[i]= decode(codebook, index[i])*
   *pseudo_index=random_number[ index[i]]*
   *swap(codebook[index_table[i]],codebook[pseudo_index])*
   *end for*

The corresponding image block is placed over the received index. This process is continued until all received indices are processed and the original image is reconstructed.

## 4. EXPERIMENTAL RESULTS

To evaluate the performance of the proposed image cryptosystems are described here. Several image codebook generation algorithms [14-16] are available in literature. However, in this paper LBG codebook generation algorithm is implemented. The codebook and image block sizes are 256 and *4 × 4* pixels respectively. As a representative, only the results of Lena and peppers image of size 512 × 512 are shown. The original images of Lena and Peppers are presented in Fig. 6.

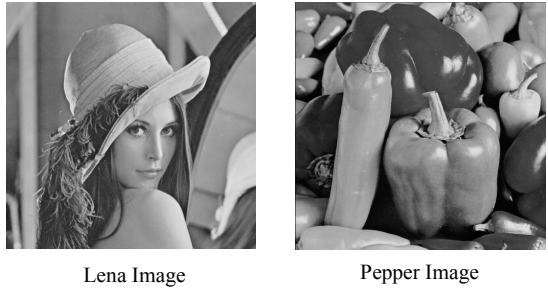

Lena Image      Pepper Image

*Fig.6.* Original Test Images

Initially random indexing of codebook is considered for encryption. But using this technique the generated index-matrix and codebook do not provide sufficient confidentiality to the image as because reconstructed images are still traceable without using pseudorandom sequence. The reconstructed image with those compressed data set are shown in *Fig. 7*. But the proposed simultaneous compression-encryption algorithm provides the expected performance because it generates shuffled codebook and index-matrix. The reconstructed image with those compressed data set are shown in Fig. 8.

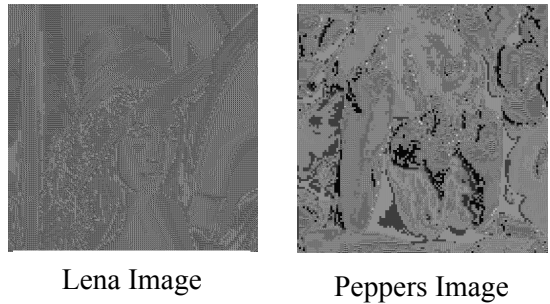

Lena Image      Peppers Image

*Fig. 7.* Reconstructed Lena and Peppers Image under considering only random indexing of the codebook

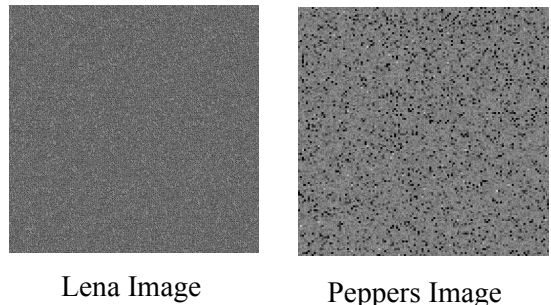

Lena Image      Peppers Image

*Fig.8.* Reconstructed Lena and Peppers Image under considering proposed simultaneous compression-encryption algorithm

## 5. SECURITY ANALYSIS

In our proposed technique compressed image components are not directly encrypted by any existing cryptographic technique like DES, RSA, RC4 and so on. The proposed technique generates shuffled index-matrix and codebook as described in section 4 to achieve the confidentiality of the input image. In this proposed scheme, we need a fixed sequence of random numbers during encoding/decoding process. Note that same PRNG and seed must be used and need to be communicated securely between the sending and receiving ends; otherwise encryption and decryption of the images cannot be done accurately. For secure transmission of seed, several key-agreement protocols are available such as Diffie-Hellman (DH) [13], Elliptic-Curve (CC), GDH.X, TGDH, CCEGK etc and one of them can be used for seed exchange. Here the seed may be considered as a key in our cryptosystem.

To prove the feasibility of our image cryptosystem, we consider its security under the following two types of attacks: Cipherimage-only attack and known-plainimage attack. Under Cipherimage-only attack the opponent is assumed to obtain information from networks, but do not have the secret key/seed. In this category there are two methods to attack. One is to directly cryptanalyze the existing cryptosystems. To do this, the cryptanalysts need to know the seed, which one is shared by key-agreement protocols such as DH. However DH key agreement gets its security from the computational difficulty of computing discrete logarithms in a finite field; hence cryptanalysts cannot obtain compressed dataset. However, the opponent tries to guess the seed by brute force technique. If the seed size is 64 bits then it has $2^{64}$ possible combinations. If the opponent uses a 1000 MIPS computer to guess the seed, the computational cost is then,

$$\frac{2^{64}}{1000 \times 10^6 \times 3600 \times 24 \times 365} > 500 \text{ years}$$

This time cost is large enough that no one image can be closed-door after 500 years. Furthermore, according to the structure of our proposed cryptosystem, if opponent try to guess the pseudo random number sequence for codebook. Then the total number of permutation of proposed cryptosystem for 256 codewords is 256!. So the sequence of pseudo random number guessing is more difficult than guessing the seed by brute force technique.

Under a known-plainimage attack, the opponent is assumed to have obtained several plainimage and cipher components pairs. In this case, the cryptanalysts intend to obtain the secret key by analyzing these pairs. After recovering the secret key, the opponent can correctly decrypt the next cipher image component if the original image is encrypted by same seed/key. To stop such type of attack, if we choose seed in such a way that the seed is changed for every image, then the scheme is well secured.

## 6. CONCLUSION

An efficient image cryptosystem based on simultaneous VQ compression and shuffling of the codebook is proposed, which in turn shuffles positions of the image blocks in the index matrix, thus provides the secrecy of them. It needs the generation of a pseudorandom sequence and for which the MATLAB operator has been used. Although it is considered to be publicly known, the overall secrecy is maintained by keeping the corresponding seed as secret to the opponents. The proposed scheme is tested in some several test images and satisfactory results have been found. It is secured also as evident from the security analysis of the proposed scheme.

About the authors

**Arup Kumar Pal** is a Junior Research Fellow in the department of Computer Science and Engineering, Indian School of Mines, Dhanbad-826004, India (phone: +91-9470958512; e-mail: arupkrpal@gmail.com).

**S. Mukhopadhyay** is an Assistant Professor in the Department of Computer Science & Engineering, Indian School of Mines, Dhanbad-826004, India
(e-mail: msushanta2001@yahoo.com).

**G. P. Biswas** is an Associate Professor in the Department of Computer Science & Engineering, Indian School of Mines, Dhanbad-826004, India     (e-mail: gp_biswas@yahoo.co.in).